# Sub-wavelength terahertz beam profiling of a THz source via an all-optical knife-edge technique

Sze Phing Ho,[1,2] Anna Mazhorova,[1] Mostafa Shalaby,[1] Marco Peccianti,[1,3*] Matteo Clerici,[1,4*] Alessia Pasquazi,[1,3] Yavuz Ozturk,[1,5] Jalil Ali,[2] and Roberto Morandotti[1]

[1]INRS-EMT, University of Quebec, 1650, Blvd. Lionel-Boulet, Varennes, Quebec J3X 1S2, Canada

[2]Institute of Advance Photonics Science, Nanotechnology Research Alliance, Universiti Teknologi Malaysia, 81310 UTM Skudai, Johor, Malaysia

[3]Department of Physics and Astronomy, Pevensey Building II, 3A8, University of Sussex, Brighton BN1 9QH, UK

[4]School of Engineering and Physical Sciences, Heriot-Watt University, Edinburgh, United Kingdom

[5]Electrical and Electronics Engineering Department, Ege University, Izmir 35100, Turkey

* m.peccianti@gmail.com, matteo.clerici@gmail.com.

Recent technological breakthroughs in the generation, manipulation and detection of long-wavelength radiation have led to the rapid development of terahertz (THz) science, with a strong boost in material analysis and characterization. This success is largely due to the nature of the field-matter interaction induced by THz radiation. Several materials of interest including explosives, illicit drugs, pollutants, biologic samples and medicines have unique spectral fingerprints over a very broad spectrum at THz frequencies, as a consequence of either molecular or structural resonances. Such information is now accessible thanks to the several recent advancements in key technologies and methodologies[1-7]. Furthermore, the low photon energy of THz radiation makes it of extreme interest for the imaging of biological samples, since it does not carry the hazard issues typically related to ionizing fields[8].

In 1995[9], Hu and Nuss pioneered THz imaging via the use of ultra-short THz pulses. Their work stimulated the exploration of novel imaging concepts relying on the time-domain discrimination of the THz electric field[10,11]. Such THz imaging systems are able to collect the three-dimensional distribution of the field, with a working principle recalling that of ultrasound medical machines. In order to provide an imaging system sensitive to sub-millimetre details, as required by several applications such as biological imaging, resolutions largely exceeding the limitations imposed by standard far-field-based imaging approaches (e.g. 300 μm at 1 THz) are commonly required. For this reason, several near-field techniques have been implemented to obtain deeply sub-wavelength (sub-$\lambda$) resolutions (even below $\lambda/100$)[12-16]. A typical approach relies on a mechanical sub-$\lambda$ probe, such as a tip or an aperture, able to probe the near-field of the THz field impinging on the sample. A possible alternative for the characterization of sub-$\lambda$ THz features relies on the implementation of a knife-edge (KE) measurement of the field radiating from the surface of the sample under investigation[17,18]. The KE technique involves the partial clipping of a propagating field induced by a thin shield with a flat boundary, i.e. a blade. In the standard setting, a power meter collects the transmitted power. The blade is then translated and the correlation between the blade position and the recorded power enables the reconstruction of the field profile. The KE technique has been successfully implemented to characterize the profiles of a variety of THz sources. In particular, we recently highlighted that the reconstruction of sub-$\lambda$ sources using KE techniques is in general characterized by severe image aberrations due to the non-separable space-time nature of the radiated non-paraxial field. Yet, exploiting the inherent field sensitivity of time-domain THz detection techniques, we demonstrated that the exact sub-$\lambda$ reconstruction of the field profiles is indeed feasible[18]. For a proper sub-$\lambda$ characterization via KE, a physical blade must be located and translated in close proximity (at a sub-$\lambda$) distance from the sample, a quite unpractical constraint in many real-world scenarios.

In this work, we propose an all-optical KE characterization technique and we demonstrate its working principle by characterizing the sub-$\lambda$ features of a spatially modulated THz source directly on the nonlinear crystal employed for the THz generation. In our experiment an ultraviolet (UV) optical beam is projected on the output facet of a generation crystal (featuring sub-$\lambda$ thickness) where it induces a thin layer (< 100 nm) of photo-excited carriers. Such conductive mask blocks THz radiation, thus acting as the blade in a KE measurement, directly on the THz source plane. We then use this approach to implement a KE measurement enabled by the optically induced virtual blade. This optical KE (OKE) technique eliminates the need of a moving physical blade close to the sample. It is also worth noticing that the OKE enables us to characterize the emission geometry *inside the generation crystal*, overcoming the significant refractive index mismatch between the crystal and the air, well known for filtering out the high frequency components of the field spatial spectrum. We employ this approach to demonstrate the imaging of a THz source with sub-$\lambda$ features (< 30 μm) directly in its emission plane.

**Results**

The experimental setup used for the demonstration of the OKE approach is shown in Fig. 1a. Terahertz pulses are generated via optical rectification in a 20 μm-thick, free standing, <110>-cut Zinc Telluride (ZnTe) crystal. The THz pulse is detected by electro-optics sampling in a 3mm-thick, <110>-cut ZnTe crystal at the Fourier plane using a parabolic mirror-based system. For the sub-$\lambda$ characterization by the KE technique it is indeed essential to detect the central part of the spatial Fourier transform of the investigated field pattern[18] (note that this configuration differs from the standard setup in THz spectroscopic systems). The virtual blade on the output facet of the generation crystal is formed by UV femtosecond pulses ($\lambda$ = 400 nm, having a photon energy of 3.1 eV) that photo-excite the carriers into the conduction band of the ZnTe semiconductor in a sharp blade-shaped area projected by a telescope system (lenses L1 and L2 in Fig. 2) to form the desired sharpness and beam waist (the beam waist is determined to be = 3mm through a conventional KE measurement, i.e. performed by translating a physical blade to block the UV beam -in the absence of the ZnTe generation crystal - at the THz generation plane). OKE measurement is implemented by translating a physical metallic blade before the telescope in the $x_0$-direction, which controls the boundary position of the optical-induced conductive layer, the virtual knife-edge, on the crystal surface.

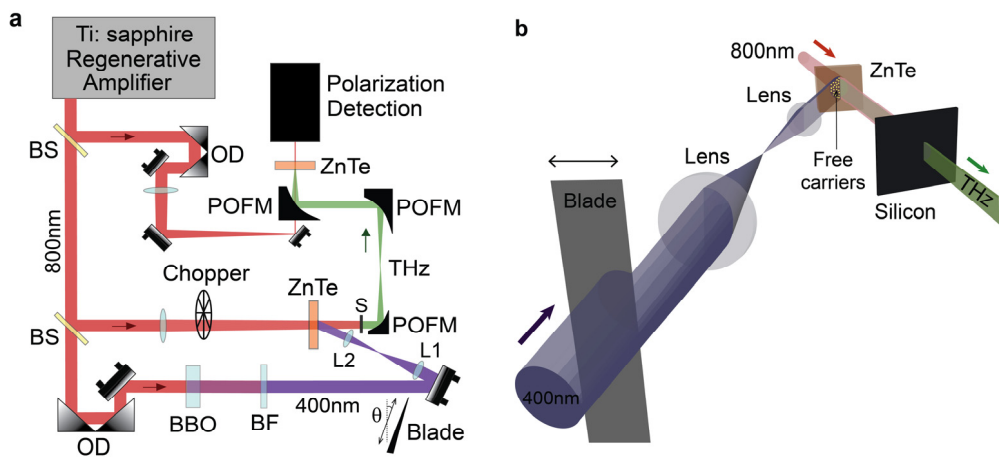

**Figure 1| OKE measurement on the THz source.** (**a**) Experimental setup (BS: beam splitter, OD: optical delay, BF: bandpass filter, POFM: parabolic off-axis mirror, L1 & L2: lenses, S: silicon, WP: Wollaston prism). (**b**) Free carriers induced by UV pulses in a blade-shaped area.

**Temporal control of the OKE technique.** The proposed OKE technique is firstly demonstrated on a super-$\lambda$ THz source. Figure 2 illustrates the detected THz wave peak for the ZnTe surface pumped by the UV radiation with fluence ≈ 330 µJ/cm$^2$, corresponding to a UV pump energy of 95 µJ, chosen to prevent optical damage in the ZnTe crystal while maximizing photo-carrier generation. The red curve shows the THz peak field against the delay between the UV pump and the THz pulse, $t_{pump,UV}$. Approximating the photo-excited region as a steep-wall metal layer, the dark grey plot shows the estimated exponential decay $\alpha d = \ln(A_R/A_S)$, i.e. the product between the effective layer thickness $d$ and its attenuation coefficient $\alpha$, where $A_R$ is the reference THz field and $A_S$ is the signal detected after propagation through the layer[19]. The maximum attenuation is reached for delays exceeding 6 ps, with approximately 85% of the total THz energy reflected by the free carriers. In the inset the peak field, plotted up to a maximum delay of 140 ps, highlights a carrier recombination time within the scale of 100 ps, consistent with the excitation of a very high photo-carrier population in the conduction band.

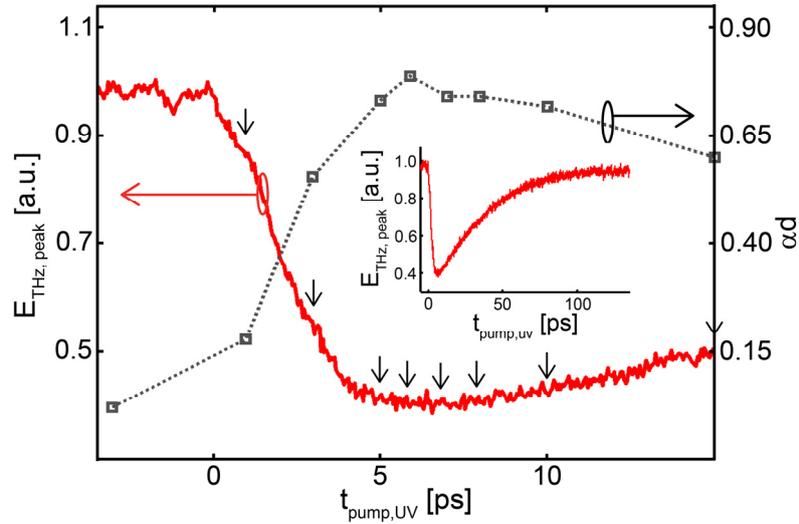

**Figure 2 | Temporal control of the OKE technique.** (Red) Peak-THz field vs. the delay $t_{pump,uv}$ between the UV pump pulse and the THz pulse (for positive values the UV pump impinges on the sample before the THz pulse). In the inset, the same plot is presented in a much larger $t_{pump,uv}$ window, highlighting the typical carrier recombination time within the 100 ps scale. In dark grey, a simple estimation of the (exponential) attenuation is calculated from the field decay. The arrows in the plot indicate the delays at which the KE measurement has been performed.

To highlight the effect of the carrier temporal dynamics, OKE measurements have been taken at several UV pump delays ($t_{pump,UV}$ = 1 ps, 3 ps, 5 ps, 6 ps, 7 ps, 8 ps, 10 ps and 15 ps), indicated by arrows in Fig. 2. Our experimental results at each pump delay are presented in Fig. 3. The set of data plotted in Fig. 3a illustrates the dependence of the THz decay from the UV pump energy. The OKE measurement performed for a UV pump energy of 95 µJ is shown in Fig. 3b where the THz power obtained by the integral of the squared field over time is plotted versus the blade position. Noteworthy, as the THz beam waist is significantly super-$\lambda$, the OKE does not introduce any significant changes in the field phase[18] (i.e. the waveform just scales as the beam is clipped by the edge). This is also highlighted in Fig. 3c, which shows the OKE measurements resolved around the frequency 1 THz ($\lambda = 300$ μm), extracted as the square-root of the THz power spectral density at 1 THz: the curve matches closely the one obtained from the full spectrum, at any tested delay.

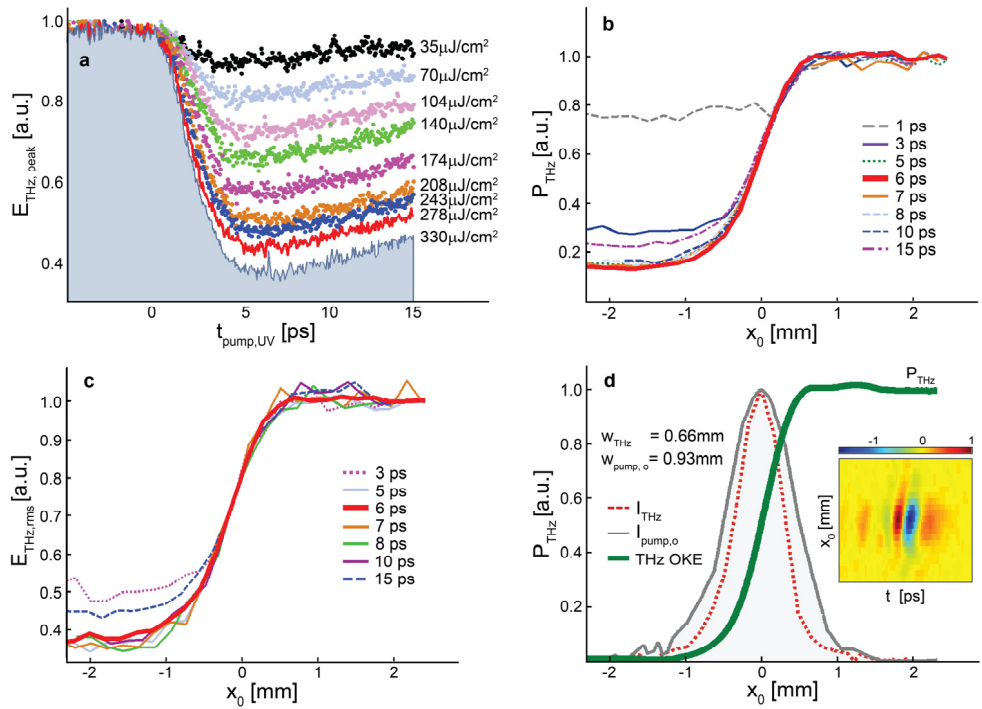

**Figure 3 | OKE results on super-$\lambda$ THz source.** (**a**) THz peak field vs. UV pump delay at different excitation fluencies. (**b**) The THz power vs. the blade position. (**c**) The OKE measurements resolved at 1 THz are plotted as the THz field versus the blade position. (**d**)

The OKE measurement for a UV pump delay of 6 ps compared with the beam waist of the optical pump. The inset shows the reconstructed spatio-temporal map.

The spatio-temporal reconstruction of the THz field, $E_{THz}(x_0, y_0, t)$, is thus obtained at a pump delay $t_{pump,uv}$ = 6 ps, under the common hypothesis of a completely separable $x$ and $y$ dependence of the field profile (i.e. $E_{THz}(x_0, y_0, t) = a(x_0, t) b(y_0, t)$). The temporal THz waveform collected by the TDS system for each blade delay provides a spatio-temporal map, $EM(x_0, t)$. The THz field, $E_{THz}(x_0, t)$, is evaluated by performing a spatial derivative of the map $EM(x_0, t)$, as reported in the inset of Fig. 3d. The THz intensity profile is depicted by the red dotted line in Fig. 3d. The waist of the THz intensity profile, i.e. in arbitrary units $I_{THz}(x_0) = \int_{-\infty}^{\infty} |E_{THz}(t)|^2 dt$ (defined as the power at $1/e^2$ of the peak), is found to be $w_{THz}$ = 0.66 mm, in perfect agreement with the estimated waist $w_{pump,o}$ = 0.93 mm of the Gaussian pump[20]. In order to confirm the validity of the estimated waist $w_{pump,o}$, we compared our results with those determined via the conventional KE technique (obtained by blocking the optical pump at the THz generation plane in the absence of the ZnTe crystal), indeed finding a very good agreement.

Given the fact that *we are mapping the THz source inside the generation crystals* –one of the main targets in this paper-, it is important to highlight that among several crystals that are mid-to-large band-gap nonlinear semiconductors, ZnTe is by far the most popular and widely deployed for THz generation due to the very favourable phase matching condition (occurring around the Titanium:Sapphire laser central frequency, i.e. at 800 nm). Thus, this approach has a significantly large applicability in the field (this may include also semiconductor-based THz generating devices such as quantum cascade lasers (QCLs)).

**Sub-$\lambda$ beam profiling.** We chiefly investigated the OKE as a mean for the characterization of sub-$\lambda$ THz features, and specifically at the THz source plane. A pump illumination pattern consisting of narrow bright fringes is generated by a Fresnel biprism made of BK7, featured by a refractive index of 1.51, a refractive angle of 1.5° and an apex angle of 177°. In order to generate the fringes, the biprism is mounted in the THz pump beam path, before the ZnTe generation crystal. The fringe spacing, $d_p$, originating from the field interference is simply determined by geometrical optics as $d_P = \lambda/[2(n-1)\theta_r] = 30$ μm, where $\lambda$ is the 800 nm pump wavelength, $n$ is refractive index of the biprism and $\theta_r$ is the refractive angle, as shown in Fig. 4a. Figure 4b-c illustrate the OKE measurement applied on the sub-$\lambda$ modulated THz source and its reconstructed spatio-temporal profile with blade movement steps of 5 μm. Figure 4c shows a better resolved acquisition performed using 1 μm blade position steps ranging from 0.5 to 0.7 mm.

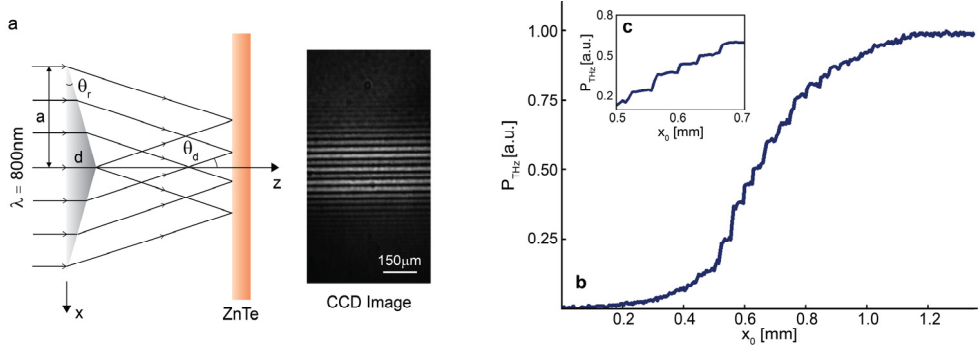

**Figure 4 | Validation of the OKE technique for the sub-$\lambda$ THz source characterization.**
(**a**) Camera shot of the THz pump profile, consisting of sub-$\lambda$ fringes of period 30 μm generated by the Fresnel biprism (see relevant details in **Methods**). (**b**) Characterization of the sub-$\lambda$ THz fringes using the proposed OKE technique. (**c**) A refined measurement taken with 1 μm blade movement steps.

Figure 5a reports the reconstructed THz spatio-temporal field $E_{THz}(x_0, t)$ evaluated as the spatial derivative of the map, EM ($x_0$, $t$) and extracted by the time resolved OKE experimental measurement on the sub-$\lambda$ THz source. A strong oscillation of the field along $x_0$ is observed if compared to a super-$\lambda$ source. In particular, one of the examples of such oscillation is presented in the upper inset of Fig. 5a, where the field oscillation at $t = 0.5$ ps is plotted. It shows that the THz fringes steps are resolved with an estimated resolution in the scale of 10 μm, which is indeed matching the diffraction limit of the telescope used to project the blade when a 400 nm pump is used. This result is compared (see Fig. 5b) with the numerical prediction of the expected THz field distribution at the clipping section for a given optical pump profile.

Noteworthy, it can be appreciated that the typical advantage of the coherent KE technique is the significantly sensitivity boosts at high transverse wavevector values. Hence this approach strongly enhances sub-$\lambda$ image features. This is clearly expressed in the comparison between the spatio-temporal spectrum extracted from the experiments (Fig. 5c) and from the calculated THz source distribution (Fig. 5d - see **Methods** for details). In the experiment, optical blade translation increments of 5 μm were considered. A weak sub-$\lambda$ modulation over the THz pulse is expected on the top of a super-$\lambda$ profile (Fig. 5b). The diffraction limit imposes that such modulation rapidly disappears in the THz propagation coordinate as it can be captured only by clipping the field at sub-$\lambda$ distances from the generation plane. This experimentally demonstrates (also in agreement with the impulsive response for point-like sources presented in Ref. 18) the inherent boost in terms of contrast associated to our method. Noteworthy, this is the first example, to the best of our knowledge, in which a THz source *has been mapped inside the generating crystal*, as the clipping of the field physically occurs before the output interface, normally characterized by a very large refractive index mismatch between the air and the crystal at THz frequencies.

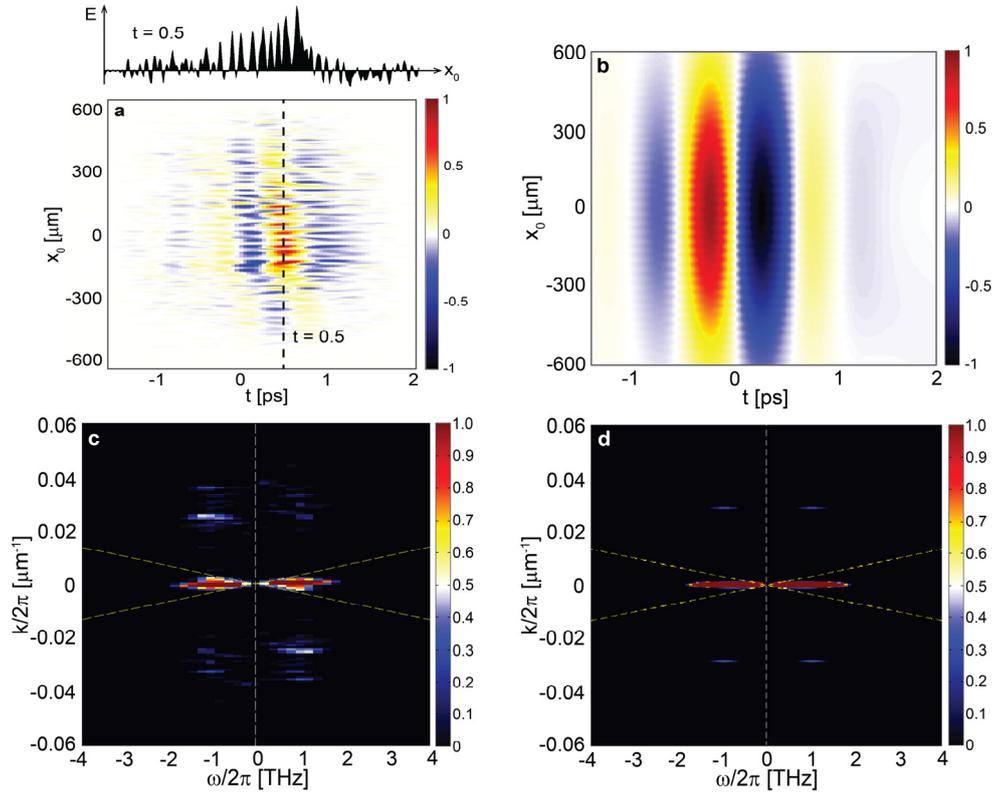

**Figure 5 | THz field retrieved in space-time and in the Fourier transformed space. (a)** Reconstructed spatio-temporal profile of the experimentally investigated grating-patterned sub-$\lambda$ THz source. The upper inset shows the field oscillation at $t = 0.5$ ps along $x_0$. **(b)** The numerically calculated spatio-temporal distribution of the source. **(c)** Experimental spatio-temporal spectra obtained from Fig. 5a in the spatio-temporal Fourier's space and **(d)** spatio-temporal spectrum extracted from Fig. 5b.

**Discussion**

The reconstructed THz pulse in Fig. 5a shows a very strong contribution of the sub-$\lambda$ fringes, indicating that the data are collected in proximity of the generation plane. This is particularly evident when the reconstructed field in the transformed space $E^{(R)}(k, \omega)$ is observed as in Fig. 5c, along with the calculated $E^{(i)}(k, \omega)$ in Fig. 5d (field incident on the blade), see **Methods** for details. In the diagrams, we report the loci $k = \pm\omega/c$, which discriminate the super-$\lambda$ ($|k| < |\omega|/c$) from the sub-$\lambda$ regime ($|k| > |\omega|/c$). The strongly visible sub-$\lambda$ components (corresponding to $k/2\pi = 0.025$ μm$^{-1}$ and $\omega/2\pi = 1$ THz) carry the information on the periodic modulation featuring the THz spatial profile. Noteworthy, there is a strong asymmetry in the reconstructed THz field profile and a much stronger contribution of the sub-$\lambda$ fringe features with respect to those calculated by a numerical model accounting for the sub-$\lambda$ THz modulation. This enhanced sensitivity to sub-$\lambda$ image features is expected for a KE+TDS system[18] (see **Methods** for details) that imposes a transfer function $\widetilde{T}(k_x, \omega)$ between the incident and the reconstructed fields in the Fourier transformed space:

$$\widetilde{E}^{(R)}(k_x, \omega) \propto \widetilde{T}(k_x, \omega) \widetilde{E}^{(i)}(k_x, \omega) \quad (1)$$

$$\widetilde{T}(k_x, \omega) \equiv \begin{cases} \sqrt{1 + \dfrac{ck_x}{\omega}} & \text{for } k_x > -\dfrac{\omega}{c} \\ 0 & \text{for } k_x < -\dfrac{\omega}{c} \end{cases} \quad (2)$$

where $c$ is the speed of light. Such function cuts the super luminal components corresponding to the region bounded by the straight lines $k_x = -\omega/c$ and $\omega = 0$, but strongly enhances the features in the region between $k_x = \omega/c$ and $\omega =$

0. As shown in Ref. 18, the proposed KE+TDS system is particularly suitable to detect weak sub-$\lambda$ details as far as the blade is positioned in close proximity to the sub-$\lambda$ features to be characterized. The proposed OKE technique satisfies this stringent condition and furthermore, allows the mapping of the sub-$\lambda$ field *inside the high refractive index medium* where the THz is generated or where the modulation is imposed.

We would like to stress that the results presented here are inherently different from those reported in previous works (e.g. Ref. 12), where the only similarity is basically the use of active photo carriers. The KE technique is not an aperture based imaging system, although the image is generated with a similar number of acquisition points (e.g. tomography). From the model above we can infer that, despite the more complex reconstruction approach, KE significantly outperforms aperture based imaging system when the dynamics and the signal-to-noise ratio (SNR) of the acquisition are limited (e.g. as in the case of THz-TDS). The signal in aperture based imaging systems (see for example, Ref. 12) is constrained by the Bethe's transmission relation for the total radiation $S_{tot} \propto r^6/\lambda^4$ reported for a circular aperture[14]. As a result, the spatial resolution of THz imaging systems is mostly limited by the SNR. In this work we experimentally observe, for the first time, that the KE response function enhances sub-$\lambda$ details with respect to their super-$\lambda$ counterparts (within the accuracy of the model hypotheses). In conclusion, our results demonstrate the possibility of a novel, all-optical, ultra-thin-thickness KE as an effective and promising characterization technique for sub-$\lambda$ features in the THz spectral region. Specifically, our setting is directly applicable to THz field-profiling inside the generation crystal. We believe that this approach could readily lead, as a next step, to the realization of reconfigurable (active) metamaterials or other structures directly imprinted on the generating crystal. We foresee the usage of a spatial light modulator (SLM) to dynamically create and move the optically induced blade.

**Methods**

**Experimental details.** The experimental setup is sketched in its main elements in Fig. 1a: THz pulses are generated through optical rectification of femtosecond optical pulses[21] delivered by a Titanium:Sapphire regenerative amplifier (Spectra-Physics *Spitfire*). The laser source emits a train of 120 fs pulses with a carrier wavelength of $\lambda$ = 800 nm, at a repetition rate of 1 kHz. The input beam is split into three lines, i.e. the pump (*pump,o*), the probe and the UV lines (*pump,UV*). In the *pump,o* line, a pulse train with an average power of 200 mW illuminates a 20μm-thick free standing <110> ZnTe crystal, generating THz radiation. Noteworthy, the thickness of the generation crystal is much smaller than the lower wavelength limit of the emitted THz power spectrum (typically for a ZnTe source $\lambda_{min}$ > 200 μm at -3 dB).

A properly designed imaging system consisting of three parabolic mirrors reconstructs the Fourier's plane of the crystal emitting-surface onto a second ZnTe crystal of thickness 3 mm, exploited to implement the electro-optic detection of the THz field. Such field is sampled at the centre of the Fourier plane, thus providing the **average field profile** (*not* the power) in the source plane.

Counter propagating UV femtosecond pulses ($\lambda = 400$ nm, of photon energy 3.1 eV) from the pump are exploited to induce a photo-carrier layer via single photon absorption at the output surface of the THz-emitting ZnTe crystal, (ZnTe is a semiconductor with a band gap of 2.3 eV at 300 K[22]). The UV pulses are generated through a type-I frequency doubling process for the 800 nm radiation in a 0.5 mm-thick BBO crystal. A bandpass filter centred at 400 nm isolates the second harmonic component. An imaging system consisting of the lenses L1 and L2 projects the image of a blade placed in the UV beam path onto the THz generation crystal. The position of the boundary of the optical-induced conductive layer (Fig. 1b), i.e. the knife-edge, is controlled by translating the blade. This thin conductive mask has a thickness orders of magnitudes smaller that the THz wavelength, below hundred nanometers (i.e. in the scale of the skin-depth of ZnTe at 400 nm) and is positioned at a sub-$\lambda$ distance with respect to the THz generation sections located within the crystal volume. Both of these conditions are required for the

sub-$\lambda$ profiling of the THz field[18]. The synchronization between the free carriers excited by the pump and the generated THz pulse is implemented via a tunable delay on the UV pump line. The THz time-domain waveforms are collected as the KE moves along the $x_0$-coordinate by translating the real blade. The experiments were performed under a dry nitrogen atmosphere in order to eliminate the typical THz signal induced by water vapour.

**Grating-patterned (rectangular-shaped) sub-$\lambda$ THz source.** A thin biprism with refractive index $n$, refraction angle $\theta_r$ and maximum thickness $d$, as shown in Fig. 4a, forms fringes when illuminated by coherent radiation. In the small angle approximation the intensity pattern after the biprism (at distance z) reads as[23]:

$$I(x,y,z) = 2\{1 + \cos[2k(n-1)\theta_r x]\} \qquad (3)$$

where $k = 2\pi/\lambda$ is the propagation constant. From simple algebra the fringes spacing results:

$$d_p = \frac{\lambda}{2\theta_d} = \frac{\lambda}{2(n-1)\theta_r} \qquad (4)$$

where $\theta_d$ is the deflection angle. In our experiment, a Fresnel biprism is mounted prior to the ZnTe crystal, and non-diffractive fringes are projected onto the ZnTe to generate a sharply sub-$\lambda$ modulated THz source.

**Numerical calculation of the THz source distribution.** The optical pump is modelled as a one dimensional source distribution along $x$ in the plane $z = 0$, generating THz radiation by optical rectification. In the hypothesis of no-depletion, the optical source intensity $I_o(x, z, \omega)$ acts as an equivalent current source in the Helmholtz equation for the THz electric field. In our experiment the optical field is polarized along $y$. By computing the nonlinear second order tensor we calculate that the equivalent current source is polarized along $y$, together with the generated electric field, obtained from the equation:

$$\nabla^2 \mathrm{E}^{(i)} + k_0^2 \mathrm{E}^{(i)} = \mathrm{J} \propto i\omega I_o(x,z,\omega). \tag{5}$$

The electric field is calculated by solving equation (7) with the two dimensional green integral[24].

**KE+TDS system transfer function.** The THz electric field is defined in time, $t$, and in frequency, $\omega$, by the Fourier relation:

$$\mathbf{e}(\mathbf{r},t) = 2\,\mathrm{Re}\int_{-\infty}^{\infty} \mathbf{E}(\mathbf{r},\omega)e^{-i\omega t}\frac{d\omega}{2\pi} \tag{6}$$

Considering $z$ the propagation coordinate, a spatial Fourier transform for the transverse coordinates $x, y$ is defined as:

$$\widetilde{\mathbf{E}}(k_x, k_y, z, \omega) = \int\int_{-\infty}^{\infty} \mathbf{E}(x,y,z,\omega)e^{-ik_x x - ik_y y}\,dx\,dy \tag{7}$$

For simplicity we consider an incident field distribution $\mathrm{E}^{(i)}(x,\omega)\hat{y}$ that is invariant and polarized along $y$, i.e. parallel to the blade edge. The blade is in the plane $z = 0$. At a coordinate $z = 0^+$, the field transmitted by the blade can be calculated via the Sommerfeld integral[24]:

$$\breve{E}(k_x, z = 0^+, \omega) = \breve{E}^{(i)}(k_x, \omega) + 2i \int \frac{\sqrt{k_0 + s_x}}{\sqrt{k_0 + k_x}} \breve{E}^{(i)}(s_x, \omega) \frac{e^{i(s_x - k_x)x_o}}{k_x - s_x} \frac{ds_x}{2\pi} \quad (8)$$

For each position of the blade moving along $x$, the TDS system implements the integral of the transmitted field along $x$ and $y$ for a specific polarization, i.e. for a field polarized along $y$ we collect the quantity:

$$\iint_{-\infty}^{\infty} \mathbf{E}(x, z = 0^+, \omega) \cdot \hat{y} \, dxdy .$$

Or equivalently (i.e. in the Fourier space), the TDS samples the function in $k_x = 0$: $\widetilde{\mathbf{E}}(k_x = 0, z = 0^+, \omega) \cdot \hat{y}$. Combining this result with the Sommerfeld integral, we find that the map extracted by the KE+TDS in function of the blade position $x_0$ is:

$$\widetilde{E}(k_x = 0, \omega) = \widetilde{E}^{(i)}(k_x = 0, \omega) - \int_{-\infty}^{\infty} \frac{2i}{s_x} \sqrt{1 + \frac{cs_x}{\omega}} \breve{E}^{(i)}(s_x, \omega) e^{is_x x_0} \frac{ds_x}{2\pi} \quad (9)$$

Deriving along $x_0$ and transforming back in the temporal domain we get the reconstructed field:

$$\begin{aligned} e_R(x_0, t) &\propto \mathbf{Re} \iint_{-\infty}^{\infty} \sqrt{1 + \frac{cs_x}{\omega}} \breve{E}^{(i)}(s_x, \omega) e^{is_x x_0 - i\omega t} \frac{ds_x d\omega}{(2\pi)^2} \\ &= \iint_{-\infty}^{\infty} \widetilde{T}(k_x, \omega) \breve{E}^{(i)}(k_x, \omega) e^{ik_x x_0 - i\omega t} \frac{dk_x d\omega}{(2\pi)^2} \end{aligned} \quad (10)$$

That leads to the transfer function:

$$\breve{T}(k_x,\omega) \equiv \begin{cases} \sqrt{1+\dfrac{ck_x}{\omega}} & \text{for} \quad k_x > -\dfrac{\omega}{c} \\ 0 & \text{for} \quad k_x < -\dfrac{\omega}{c} \end{cases} \qquad (11)$$

**Acknowledgments**

The work has been supported by the MERST (Ministère de l'Enseignement supérieur, de la Recherche et de la Science) and by the NSERC (National Science and Engineering Research Council) in Canada. Ho Sze Phing gratefully acknowledges a fellowship from the Skim Latihan Akademik IPTA (SLAI) from the Ministry of Higher Education Malaysia (MOHE). Mostafa Shalaby would like to acknowledge the financial support from "Le Fonds Québécois de la Recherche sur la Nature et les Technologies" (FQRNT). Marco Peccianti acknowledges the support from the FP7 Marie Curie Actions of the European Commission, via the Career-Integration Grant (contract-Nº 630833). Matteo Clerici acknowledges the support from the FP7 Marie Curie Actions of the European Commission, via the International-Outgoing-Fellowship (contract-Nº 299522). Alessia Pasquazi acknowledges the support from the FP7 Marie Curie Actions of the European Commission, via the International Incoming Fellowships (contract-Nº 327627).

**Author Contributions**

All the authors contributed to the realization of the work and to drafting the manuscript. MP conceived the experiment. RM supervised the project.

**Additional Information**

Competing financial interests

The authors declare no competing financial interests